\documentclass{aa}  

\newcommand{\degree}{\ensuremath{{}^{\circ}}\xspace}

\bibpunct{(}{)}{;}{a}{}{,} 
\usepackage{natbib}
\usepackage{graphicx}
\usepackage{lscape}
\usepackage{rotating}
\usepackage{subfig}
\usepackage{placeins}
\usepackage{hyperref}

\usepackage{txfonts}
\begin{document} 

   \title{New orbital ephemerides for the dipping source 4U 1323-619: constraining the distance to the source}
%
%

\author{A. F. Gambino\inst{\ref{inst1}} \and R. Iaria\inst{\ref{inst1}}\and
 T. Di Salvo\inst{\ref{inst1}}\and M. Del Santo\inst{\ref{inst2}}\and L. Burderi\inst{\ref{inst3}} \and M. Matranga\inst{\ref{inst1}}\and F. Pintore\inst{\ref{inst3}}\and A. Riggio\inst{\ref{inst3}} \and A. Sanna\inst{\ref{inst3}}}

\institute{Università degli Studi di
  Palermo, Dipartimento di Fisica e Chimica, via Archirafi 36 - 90123 Palermo, Italy\label{inst1}\\
  \email{angelofrancesco.gambino@unipa.it}
               \and
           Istituto Nazionale di Astrofisica, IASF Palermo,
           Via U. La Malfa 153, I-90146 Palermo, Italy\label{inst2}
                   \and
                   Università degli Studi di Cagliari, Dipartimento di Fisica, SP
           Monserrato-Sestu, KM 0.7, 09042 Monserrato, Italy\label{inst3}  
           }    

            
%


 %
  \abstract
   {4U 1323-619 is a low mass X-ray binary system that shows type I X-ray bursts and dips. The most accurate estimation of the orbital period is 2.941923(36) hrs and a distance from the source that is lower than 11 kpc has been proposed.}
   {We aim to obtain the orbital ephemeris, the orbital period of the system, as well as its derivative to compare the observed luminosity with that predicted by the theory of secular evolution. }
   {We took the advantage of about 26 years of X-ray data and grouped the selected observations when close in time. We folded the light curves and used the timing technique, obtaining 12 dip arrival times. We fit the delays of the dip arrival times both with a linear and a quadratic function.}
   {We locate 4U 1323-619 within a circular area centred at RA (J2000)= 201.6543\degree and DEC (J2000)= -62.1358\degree with an associated error of 0.0002\degree, and confirm the detection of the IR counterpart already discussed in literature. We estimate an orbital period of P=2.9419156(6) hrs compatible with the estimations that are present in the literature, but with an accuracy ten times higher. We also obtain a constraint on the orbital period derivative for the first time, estimating $\dot{P}=(8\pm 13)\times 10^{-12}$ s/s. 
Assuming that the companion star is in thermal equilibrium in the lower main sequence, and is a neutron star of 1.4 M$_{\odot}$, we infer a mass of 0.28$\pm$0.03 M$_{\odot}$ for the companion star.
Assuming a distance of 10 kpc, we obtained a luminosity of (4.3$\pm$0.5)$\times 10^{36}$ erg s$^{-1}$, which is not in agreement with what is predicted by the theory of secular evolution. Using a 3D extinction map of the K$_{s}$ radiation in our Galaxy, we obtain a distance of 4.2$^{+0.8}_{-0.7}$ kpc at 68\% confidence level. 
This distance implies a luminosity estimation of (0.8$\pm$0.3)$\times 10^{36}$ erg s$^{-1}$, which is consistent with the adopted scenario in which the companion star is in thermal equilibrium.}
   {}

\keywords{stars: neutron – stars: individual (4U 1323-619) — X-rays: binaries — X-rays: stars – Astrometry and celestial mechanics:
ephemerides}

\maketitle

\section{Introduction}


\FloatBarrier

\begin{table*}
\caption{Observation log.} \label{tab:obs_log}   
     
\centering          
\scriptsize
\begin{tabular}{cccccc}     
\hline\hline       

Point &
Satellite/Instrument &
Observation ID &
Start time  & 
Stop time  &
T$_{fold}$ \\

 &
 &
 &
(UT) & 
(UT) &
(TJD;TDB)\\

\hline

1 &     
EXOSAT/ME &
1401 &
1985 Feb 11 18:48:49 & 
1985 Feb 13 04:39:18 &
6108.48892929229 \\

2 &     
ASCA/GIS2 &
42005000  & 
1994 Aug 04  12:21:11 &
1994 Aug 04 20:10:47 &
 9568.66237498892 \\

3 &     
RXTE/PCA &
P20066-02-01-00, P20066-02-01-01, P20066-02-01-02, &
1997 Apr 25 22:04:48 &
1997 Apr 28 03:44:49  & 
 10565.04185436878\\
 
  &     
 &
 P20066-02-01-03, P20066-02-01-04 &
  & 
  &
 \\

4 &     
BeppoSAX/MECS &
20102001 &
1997 Aug 22 17:06:09  & 
1997 Aug 24 02:02:39 &
10683.398658095225\\

5 &     
RXTE/PCA &
P40040-01-01-000, P40040-01-01-00,  P40040-01-02-000, &
1999 Jan 18 02:42:30  & 
1999 Mar 13 09:10:46  &
11223.245347406 \\

 &      
 &
 P40040-01-02-00, P40040-01-03-000, P40040-01-03-00 &
  & 
  &
 \\

6 &     
ASCA/GIS2 &
47015000 &
2000 Feb 02 12:16:04 & 
2000 Feb 04 23:19:15 &
 11577.746691772565\\

7 &     
XMM/Epic-pn &
0036140201 &
2003 Jan 29 09:03:42  & 
2003 Jan 29 22:57:07  &
12668.666973091815 \\

8 &     
RXTE/PCA &
P70050-03-01-00, P70050-03-01-01, P70050-03-01-02 &
2003 Sep 25 07:54:56  & 
2003 Sep 25 23:57:20  &
12907.661355385295 \\

9 &     
RXTE/PCA &
P90062-03-01-010, P90062-03-01-01, P90062-03-01-00, &
2004 Dec 30 21:46:40  & 
2004 Dec 31 13:53:03 &
13370.240516395665 \\

 &      
 &
 P90062-03-01-02 &
  & 
  &
 \\

10 &            
Suzaku/XIS0 &
401002010 &
2007 Jan 09 11:50:53  & 
2007 Jan 10 21:58:37 &
14110.204689820065 \\

11 &            
RXTE/PCA &
P95442-01-01-00, P95442-01-01-01, P95442-01-01-02,   &
2010 Dec 25 08:01:54 & 
2011 Mar 28 12:14:06 &
15601.92228420802 \\

 &      
 &
P96405-01-01-01, P96405-01-02-00, P96405-01-02-01 &
  & 
  &
 \\

&       
 &
P96405-01-01-00, &
  & 
  &
 \\

12 &            
Chandra/HETG &
13721, 14377 &
2011 Dec 19 01:03:11 & 
2011 Dec 24 19:01:07 &
15916.91815760006 \\

\hline

\hline
              
\end{tabular}

\end{table*}


\object{4U 1323-619} is a low mass X-rays binary system (LMXB) that was discovered by \cite{Forman} and identified as persistent by \cite{Warwick}. Periodic dips and type I bursts were discovered in the European X-ray Observatory Satellite (\textit{EXOSAT}) light curves \citep{Van_der_klis, Parmar}. 
The matter transferred from the companion star, impacting onto the outer accretion disc,
forms a bulge of cold (and/or partially ionised) matter that photoelectrically absorbs part of the X-ray emission, which comes from the inner region of the system. The shape of a dip is generally irregular and varies from one cycle to
another. However, the periodic occurrence of the dips in the light curve is strictly connected to the orbital motion of the binary system and the study of their periodicity can give information on the orbital period of the system. 
From \textit{EXOSAT} and \textit{BeppoSAX} lightcurves, \cite{Parmar} and \cite{Belucinska_beppo} inferred an orbital period of 2.932(5) hrs and 2.94(2) hrs, respectively.
\cite{Levine}, studying the RossiXTE/ASM 1.5-12 keV light curve from 1996 to 2011, find an orbital period of 2.941923(36) hrs. Although the light curves of 4U 1323-619 show dips, they do not show  eclipses, which implies that the source has an inclination angle $i$ between 60\degree and 80\degree \citep{Frank}. 
\cite{Zolotukhin}, using a \textit{Chandra} observation performed in continuous clocking mode, obtain the X-ray position of the source that, however, suffers from the indetermination in one direction because of the \textit{Chandra} observational mode. The same authors also estimate  the X-ray position of the source with an associated error of 3" using data of the X-ray Multi Mirror (\textit{XMM-Newton}) mission. 
\cite{Zolotukhin}, analysing the 2MASS catalogue, identify two possible infrared counterparts of 4U 1323-619 that are associated with the X-ray position estimated using \textit{XMM-Newton} data. From the intersection of the positional error boxes of \textit{Chandra} and \textit{XMM-Newton,} they suggest that the probable infrared counterpart is a source with a magnitude of 18.12$\pm$0.20  in the K$_{s}$ band. 

The spectral analysis of the source shows that the equivalent column density of neutral hydrogen, N$_{H}$, is large and depends on the model used to fit the spectra and on the energy band covered. \cite{Belucinska} find N$_{H}=(3.2^{+0.1}_{-0.1})\times10^{22}$ cm$^{-2}$ using \textit{Suzaku} data, \cite{Boirin} infer N$_{H}=(3.5^{+0.1}_{-0.2})\times10^{22}$ cm$^{-2}$ using \textit{XMM-Newton} data and \cite{Parmar} obtained N$_{H}=(4.0\pm 0.3)\times10^{22}$ cm$^{-2}$ using \textit{EXOSAT} data. 
\cite{Smale} suggested that the large extinction to the source also explains why the optical counterpart has not been detected yet. Furthermore, \cite{Belucinska_beppo} suggested the possible presence of a local dust halo surrounding 4U 1323-619. The main effect of the halo is to absorb part of the source radiation owing to its optical depth, which depends on the inverse of the incoming radiation energy \citep{Predel}. 

The spectral analysis of 4U 1323-619 also highlights that the energy spectrum is dominated by a power-law component. \cite{Parmar} found a photon index $\Gamma$ of the power-law component of 1.53$\pm$0.07, \cite{Boirin} obtain a $\Gamma$ of $1.9^{+0.06}_{-0.10}$ and, adopting a cut-off power-law component, \cite{Belucinska} find a $\Gamma$ of $1.67^{+0.10}_{-0.03}$ and an energy cut-off of 85$^{+77}_{-35}$ keV.
\cite{Galloway}, analysing the type I X-ray bursts properties of 4U 1323-619 with data obtained with the proportional counter array (PCA) on board the \textit{RXTE} mission, only infer  an upper limit for the distance from the source because they do not find any evidence of photospheric radius expansion (PRE). The upper limit is of 11 kpc, assuming a companion star with cosmic abundances, and of 15 kpc, assuming a pure hydrogen companion star.
Furthermore, the same authors show that the light curves demonstrate many double bursts where the secondary bursts are fainter and  suggest that this  can be explained taking into account a mixed composition of hydrogen and helium of the companion star, which causes short recurrence bursts generated by the hydrogen ignition onto the neutron star.
In addition, using EXOSAT/ME data in the energy band 1.5-11 keV, \cite{Parmar} gave a lower limit of 10 kpc for the distance from the source on the basis of the observed properties of the bursts. 

\cite{Zolotukhin}, studying the photometric properties of 4U 1323-619 in the IR band and assuming a distance of 10 kpc, find a discrepancy of one order of magnitude between the observed value of the flux and that predicted by a model that describes the system as an accretion disk illuminated by a central spherical hot corona, which has a luminosity of 5.2$\times 10^{36}$ erg s$^{-1}$ in the 0.1-10 keV band \citep{Boirin}. Based on this discrepancy, they suggest a distance from the source between 4 and 5 kpc.\\

In this work, we find the first orbital linear and quadratic ephemerides of 4U 1323-619, using all the available X-ray pointing observations from 1985 to 2011. We find a weak constraint on the orbital period derivative. 
We compare the obtained results with the prediction of the secular evolution of the binary system \citep[see e.g.][]{King, Verbunt} and suggest that the source distance is less than 5 kpc (close to 4.2 kpc for a companion star in thermal equilibrium).

\section{Observation and data reduction}


\begin{table*}
\caption{Best-fit parameters obtained by fitting the dips in the folded light curves.} \label{tab:fit_par}             
     
\centering          
\scriptsize
\begin{tabular}{cccccccccc}     
\hline\hline       

Point &
Phase interval &
Y$_{1}$&
Y$_{2}$  & 
Y$_{3}$  &
$\phi_{1}$ &
$\phi_{2}$ &
$\phi_{3}$ &
$\phi_{4}$ &
$\chi^{2}_{red}$ (d.o.f.)\\

 &
 &
count s$^{-1}$&
count s$^{-1}$& 
count s$^{-1}$&
 &
 &
 &
 &
\\

\hline

1 &     
0 - 1 &
4.88$\pm$ 0.04 &
3.29$\pm$ 0.05  & 
4.94$\pm$ 0.04  &
0.248$\pm$ 0.008 &
0.322$\pm$ 0.008 &
0.519$\pm$ 0.007 &
0.604$\pm$ 0.008 &
1.18(506)\\

2 &     
0 - 1 &
1.41$\pm$0.02  & 
0.78$\pm$0.02  &
1.51$\pm$0.02 &
 0.351$\pm$0.006&
0.384$\pm$0.005 &
0.598$\pm$0.005 &
 0.621$\pm$0.006&
1.66(438) \\

3 &     
0.6 - 1.6 &
19.85$\pm$0.05 &
15.62$\pm$0.07  & 
20.32$\pm$0.06 &
0.920$\pm$0.005 &
1.009$\pm$0.004 &
1.159$\pm$0.005 &
1.301$\pm$0.005 &
5.44(482)\\

4 &     
0 - 1 &
0.954$\pm$0.008 &
0.634$\pm$0.008  & 
0.961$\pm$0.010  &
0.359$\pm$0.007&
0.432$\pm$0.006 &
0.643$\pm$0.008 &
0.744$\pm$0.009 &
1.56(506)\\

5 &     
0 - 1 &
21.21$\pm$0.05 &
16.53$\pm$0.06  & 
21.06$\pm$0.05  &
0.334$\pm$0.004 &
0.414$\pm$0.004 &
0.580$\pm$0.003 &
0.651$\pm$0.003 &
4.89(506)\\

6 &     
0 - 1 &
1.231$\pm$0.012 &
0.724$\pm$0.011  & 
1.260$\pm$0.012  &
0.294$\pm$0.008 &
0.409$\pm$0.007 &
0.615$\pm$0.007 &
0.676$\pm$0.007 &
1.50(506)\\

7 &     
0.3 - 1.3 &
28.4$\pm$0.2 &
22 (fixed)  & 
27.4$\pm$0.2  &
0.632$\pm$0.008 &
0.699$\pm$0.007 &
0.937$\pm$0.007 &
0.983$\pm$0.007 &
15.65(506)\\

8 &     
0.6 - 1.6 &
19.31 $\pm$0.08 &
14.5 (fixed)  & 
18.96$\pm$0.07  &
0.936$\pm$0.006 &
1.052$\pm$0.005 &
1.132$\pm$0.006 &
1.246$\pm$0.007 &
2.72(495)\\

9 &     
0.9 - 1.9 &
17.34$\pm$0.13 &
12.83$\pm$0.10  & 
17.62$\pm$0.12  &
1.224$\pm$0.007 &
1.259$\pm$0.008 &
1.531$\pm$0.006 &
1.589$\pm$0.006 &
4.94(495)\\

10 &            
0.3 - 1.3 &
0.922$\pm$0.009 &
0.500$\pm$0.009  & 
0.932$\pm$0.010  &
0.629$\pm$0.006 &
0.686$\pm$0.005 &
0.881$\pm$0.008 &
1.028$\pm$0.009 &
1.65(505)\\

11 &            
1 - 2 &
18.42$\pm$0.11 &
13.64$\pm$0.11  & 
18.39$\pm$0.09  &
1.278$\pm$0.006 &
1.338$\pm$0.006 &
1.537$\pm$0.008 &
1.630$\pm$0.010 &
2.09(505)\\

12 &            
0.27 - 1.27 &
0.781$\pm$0.007 &
0.494$\pm$0.005  & 
0.813$\pm$0.007  &
0.561$\pm$0.006 &
0.607$\pm$0.005 &
0.914$\pm$0.004 &
0.956$\pm$0.005 &
2.72(505)\\

\hline

\hline
              
\end{tabular}

\end{table*}

To analyse the dips arrival times of 4U 1323-619, we used all the available X-ray archival data that include \textit{RXTE}, \textit{Chandra}, \textit{XMM-Newton}, \textit{Suzaku}, \textit{BeppoSAX}, \textit{EXOSAT} and \textit{ASCA} observations. 

\begin{figure}
         \centering
         
        \includegraphics[angle=0, width=9cm]{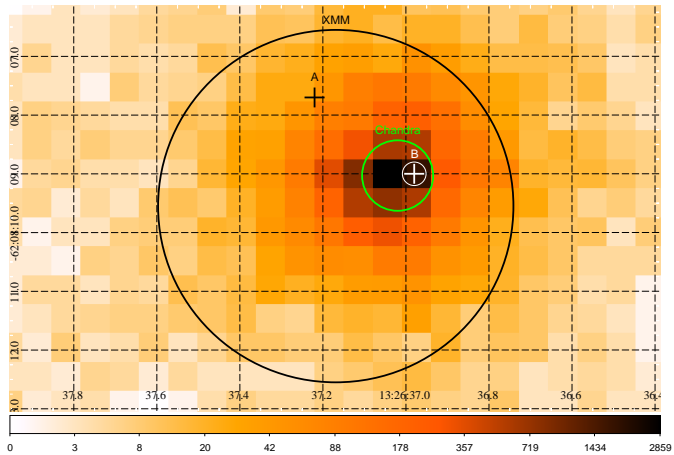}
  
         \caption{Chandra/HETG image of 4U 1323-619. The colour scale reports the number of photons. The green circle indicates the X-ray position obtained in this work, while the black circle indicates the position obtained by \cite{Zolotukhin} using \textit{XMM-Newton} data. The A and B crosses are the two candidate infrared counterparts identified by \cite{Zolotukhin}.}
         
         \label{fig:source_position}
        \end{figure}

We used the \textit{Chandra} data of the ObsIDs 13721, 14377, and 3826. Both the ObsID 13721 and ObsID 14377 data were collected in December 2011 in timed graded mode, while the data of the ObsID 3826 were taken in continuous clocking mode in September 2003.
To process the \textit{Chandra} data we used CIAO v. 4.7. The observations have been reprocessed with the \verb|chandra_repro| routine. 
We used the ObsID 13721 to obtain an accurate estimation of the X-ray source position using the \textit{Chandra} tool \verb|tg_findzo|. The revised coordinates for 4U 1323-619 are: RA (J2000)= 201.6543\degree and DEC (J2000)= -62.1358\degree, with an associated error of 0.6", which represents the 90\% confidence level \textit{Chandra} positional accuracy\footnote{See http://cxc.harvard.edu/cal/ASPECT/celmon/ for more details.}. \\
We report the Chandra/HETG image of 4U 1323-619 that was obtained from the reprocessed level 2 data in Fig. \ref{fig:source_position}. The green and black circles indicate the X-ray source position shown in this work and in \cite{Zolotukhin}, respectively. The infrared sources, A and B, suggested by \cite{Zolotukhin} are indicated with crosses and have an error circle with a radius of 0.2". The B source is located inside a circular area, which has a radius of 0.2" that has been determined with the 2MASS catalogue. 
On the basis of our results we confirm that the source B is the infrared counterpart of 4U 1323-619 as suggested by \cite{Zolotukhin}.
Hereafter we use the new X-ray \textit{Chandra} coordinates to apply the barycentric corrections.\\
We applied the barycentric corrections to the Level 2 events file using the \verb|axbary| tool and extracted the first diffraction orders and background-subtracted HEG + MEG light curves using the \verb|dmextract| tool. To subtract the background from the light curves, we selected a background events extraction region of 5" of radius, far away from the real source position. 
        
        \begin{figure}
         \centering
         
         \subfloat[]{\includegraphics[angle=-90, width=9cm]{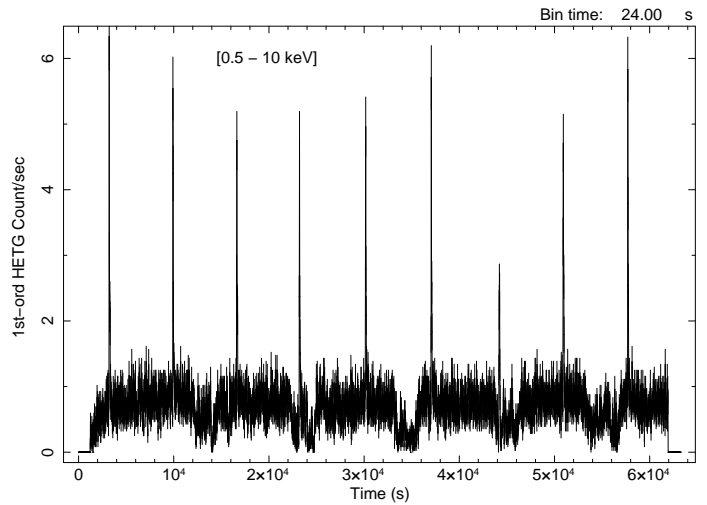}}\\
         \subfloat[]{\includegraphics[angle=-90, width=9cm]{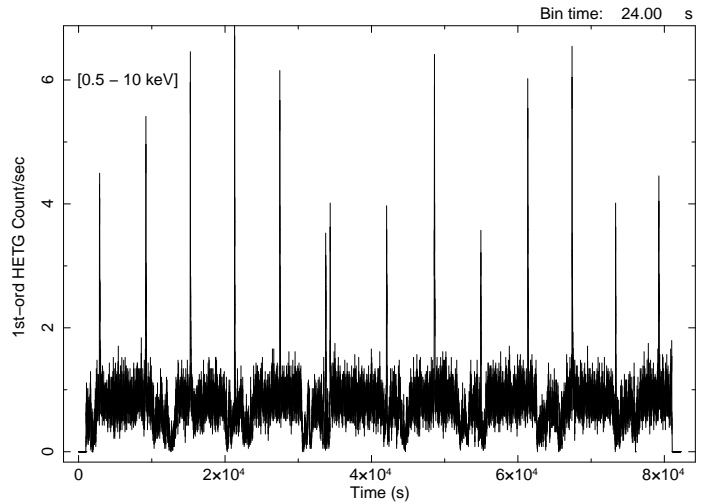}}
  
         \caption{1st order background subtracted Chandra/HETG light curves of the observations 14377 (a) and 13721 (b) in the energy range 0.5-10 keV. The light curve (a) starts on 19 December 2011 at 01:03:10 UT, while the light curve (b) starts on 23 December 2011 at 20:11:09 UT. The bin time is of 24 s.}
         
         \label{fig:curve}
        \end{figure}
The available RXTE/PCA observations span a period of about 13 years (from 1997 to 2011). We produced background-subtracted light curves, including all the energy channels and obtained the data from the HEASARC data archive. These light curves have a temporal resolution of 0.125 s and the barycentric corrections have been applied using the \verb|faxbary| tool. \\ 
The only available observation of XMM-Newton/Epic-pn is the ObsID 0036140201 of January 2003, performed in timing mode. The data have been processed with the \verb|epproc| tool of the Scientific Analysis System (SAS) v. 14.0.0 and the barycentric corrections have been applied with the \verb|barycen| tool. The extraction region of the source photons has been chosen on the basis of the analysis of the histogram showing the number of photons versus the RAWX coordinate of the image. We selected a box region with a width of 21 RAWX and centred at the RAWX coordinate of the peak of the photons' distribution (RAWX=36). Thus, we extracted the light curve with the \verb|evselect| routine, selecting events with PATTERN$\leq$4 (single and double pixel events) and FLAG=0 (to ignore spurious events), in the energy range between 0.5 and 10 keV. We binned the resulting light curve at 1 s. \\
The available \textit{Suzaku} observation of 4U 1323-619 is the  ObsID 401002010 of January 2007.
This has been processed with the \verb|aepipeline| routine and the two available data formats (3X3 and 5X5) have been unified with each other. We applied the barycentric corrections to the events file with the \verb|aebarycen| routine. We extracted the 0.2-10 keV light curve using the \verb|xselect| tool and adopting a circular region around the brightest pixel of the source with a radius of 50". We used a bin time of 16 s. \\
We also used the EXOSAT/ME observation performed on February 1985. The  light curve has been extracted in the energy range between 1 and 8 keV with a bin time of 10 s and the barycentric corrections have been applied with the ftool \verb|earth2sun|.\\
Furthermore, we used two ASCA/GIS2 observations: the first has been taken on August 1994 and the second one on February 2000.
Both  observations were performed in medium bit rate mode and the light curves were extracted with a bin time of 0.5 s. We applied the barycentric corrections using the \verb|timeconv| ftool. \\
Finally, the only available observation of \textit{BeppoSAX} is of August 1997.
The BeppoSAX/MECS light curve has been extracted with the \verb|xselect| ftool in a circular region of 4' of radius, without energy filters and with a bin time of 4 s. The barycentric corrections have been applied with the ftool \verb|earth2sun|.\\

\section{Data analysis}

\begin{figure}
         \centering
         
         \includegraphics[angle=-90, width=9cm]{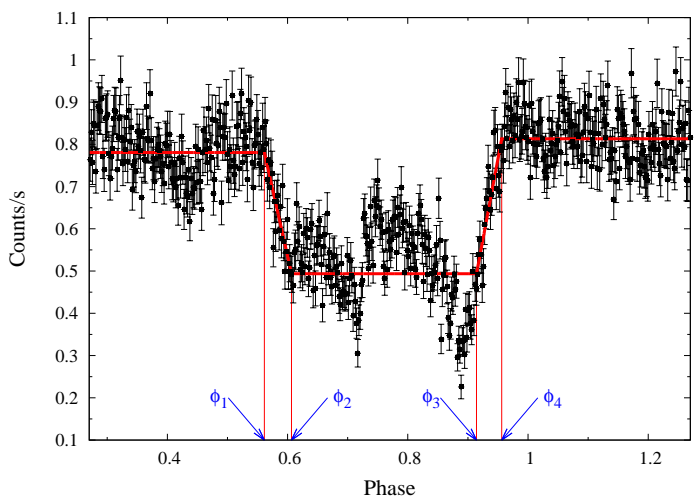}
  
         \caption{Folded light curve of the jointed Chandra observations ObsID 14377 and ObsID 13721. The red line is the step-and-ramp function that better fits the dip. The blue arrows highlight the phases of ingress ($\phi_{1}$ and $\phi_{2}$) and egress ($\phi_{3}$ and $\phi_{4}$) of the dip.}
         
         \label{fig:folded_step_ramp}

         \end{figure}
         
\begin{figure}[!h]
         \centering
         
        \includegraphics[width=9cm]{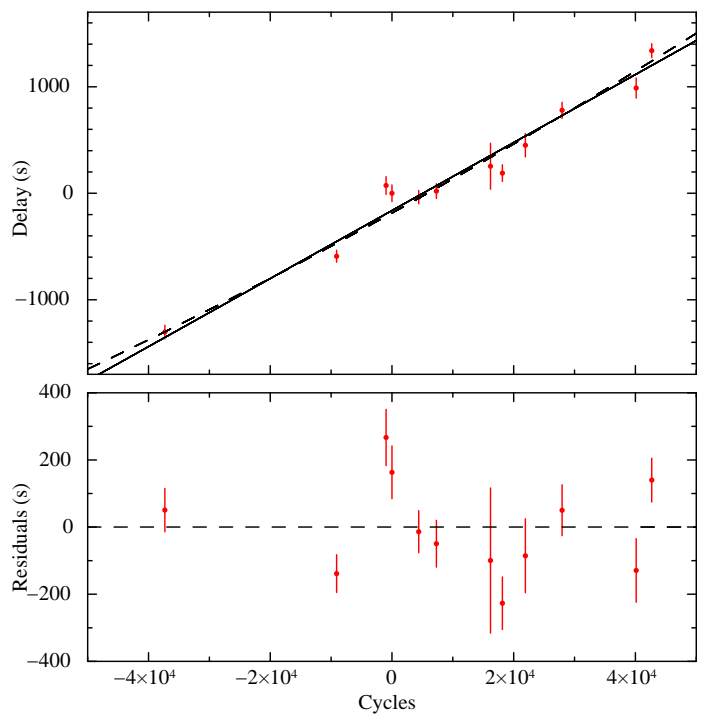}
  
         \caption{Upper panel: delays of the dip arrival times (calculated with respect to P$_{0}$ of 10590.864 s and T$_{0}$ of 10683.3987 TJD) as a function of the number of cycles.  Both the linear and the quadratic models are shown as a solid and a dashed line, respectively. Lower panel: residuals associated with respect to the linear fit.}
         
         \label{fig:fit_trend}
        \end{figure}

\begin{figure*}
         \centering
         
         \includegraphics[angle=0, width=9cm]{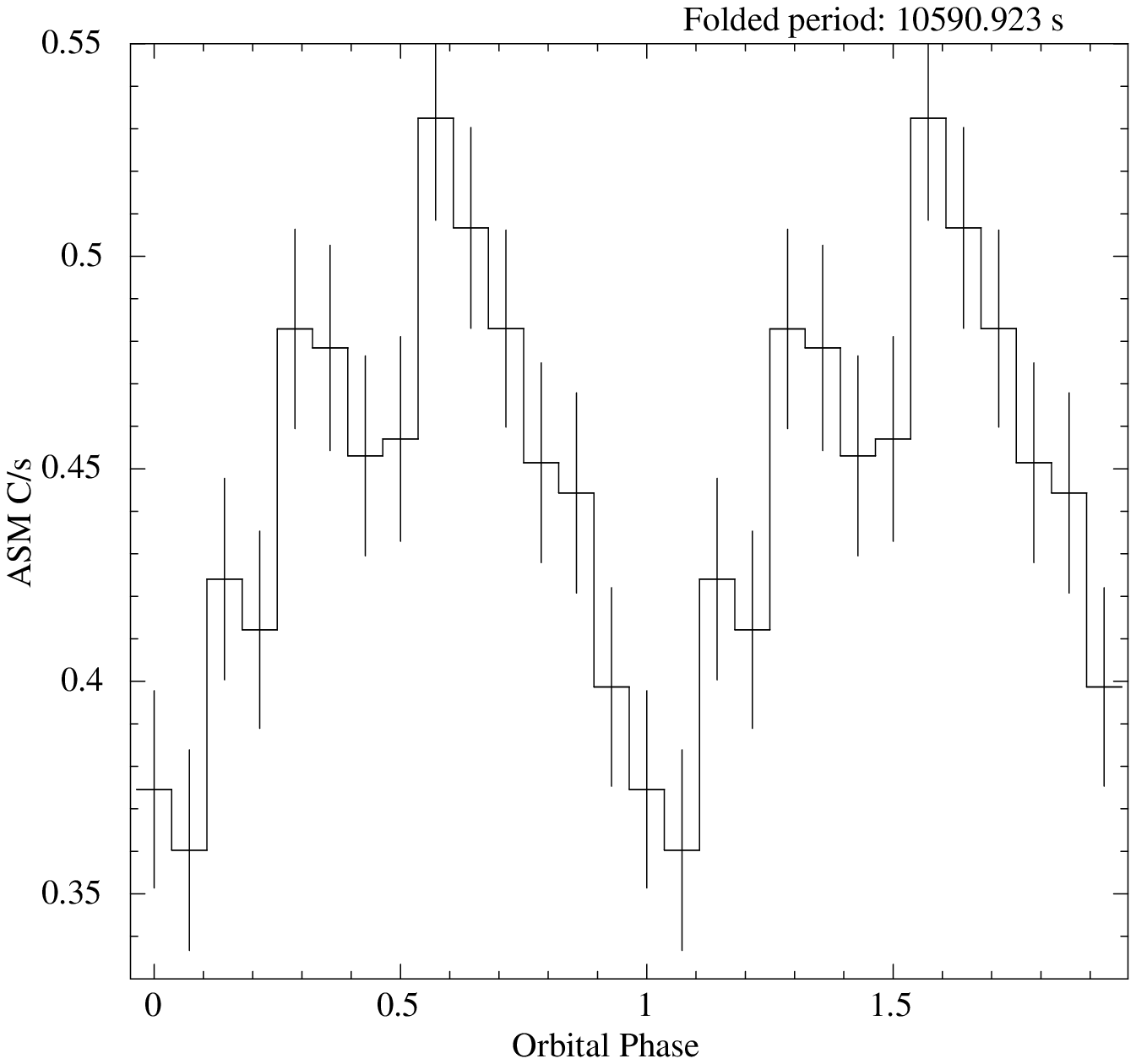}
         \includegraphics[angle=0, width=9cm]{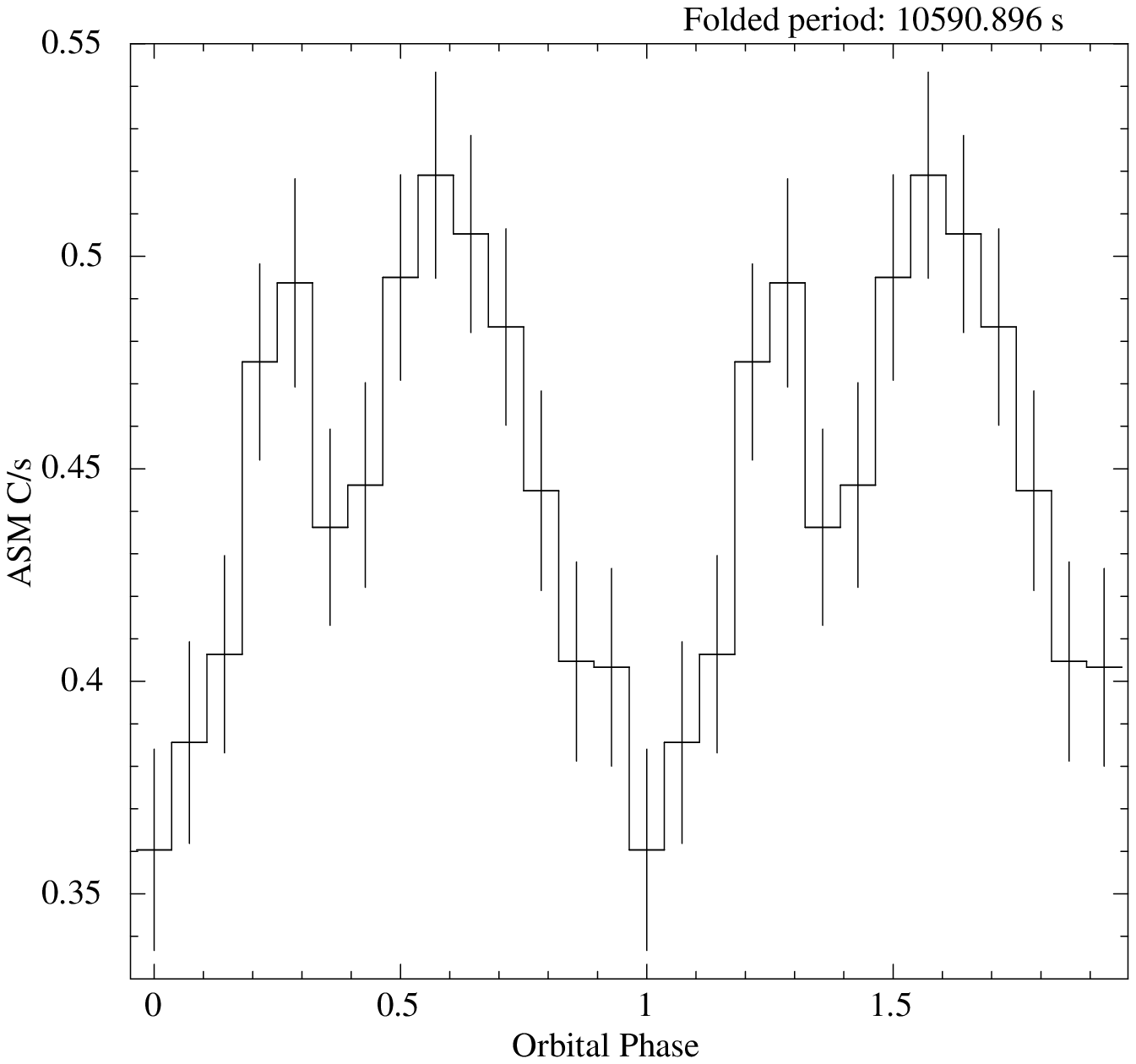}
  
         \caption{Folded RXTE/ASM light curves of 4U 1323-619 in the 2-10 keV energy band using the T$_{0}$ and orbital period suggested by \cite{Levine} (left panel) and the linear ephemeris shown in Eq. \ref{eq:linear_ephem} (right panel). We adopted 14 phase-bins per period.}
         
         \label{fig:folded_comparison}

         \end{figure*}
        
The whole analysed data-set spans a temporal range of about 26 years. 
All the selected observations have been grouped when close in time to obtain folded light curves  to increase the statistics of the dips. 
We grouped the Chandra ObsIDs 13721 and 14377. The light curves of these observations are separately shown in Fig. 2a and 2b with a bin time of 24 s. The curves have been extracted from the first diffraction order of the HETG grating and the individual observations have a duration of about 62 and 80 ks, respectively. In the ObsID 14377 five dips occur, plus one more at the beginning of the observation that is only partially visible. In addition, during the observation 13 721 seven dips occur, plus one more that is partially visible at the beginning of the observation. In both observations several Type I bursts occur. The persistent emission is mainly at about 0.8 counts/s and the count-rate at the bottom of the dips is close to 0.3 counts/s. \\
Three RXTE/PCA observations have been ruled out from our analysis: the observation whose ObsID is P40040-01-02-01 because the dip is only partially visible in the light curve, as well as the observations whose ObsIDs are P96405-01-03-00 and P96405-01-04-00, owing to the lack of dips. 
We furthermore excluded four additional RXTE/PCA observations from our analysis (ObsIDs P96405-01-05-00, P96405- 01-05-01, P96405-01-05-02, and P96405-01-05-03) and one Chandra observation (ObsID 3826) owing to the fact that their folded light curves show a high count-rate variability outside the dips that does not allow us to obtain a valid fit for the dips by using the step-and-ramp function that we define below.\\
After this preventive selection process, we analysed 12 grouped and barycentric-corrected light curves that contain a total of 95 dips. The observations that we selected for the analysis, as well as their grouping, are shown in Tab. \ref{tab:obs_log}. 
Bursts were excluded from each light curve by removing the segments starting 5 s before the rise time of each burst and ending 100 s after the peak time. We verified that the shape of the dips in each light curve is quite similar from one cycle to another and folded the light curves with a reference time and a trial period (T$_{fold}$ and P$_{0}$, respectively). To estimate the dip arrival time, we adopted the same procedure used by \cite{Iaria_2015}. 
For each light curve T$_{fold}$ is defined as the average value between the start and stop time of the observation. We fit the dip in each folded light curve with a step-and-ramp function that involves seven parameters: the count-rate before (Y$_{1}$), during (Y$_{2}$) and after (Y$_{3}$) the dip, and the phases of ingress ($\phi_{1}$ and $\phi_{2}$) and egress ($\phi_{3}$ and $\phi_{4}$) of the dip. 
We show the step-and-ramp function fitting the folded light curve of the joint Chandra observations ObsID 14377 and ObsID 13721 in Fig. \ref{fig:folded_step_ramp}.\\
The phase of the dip arrival has been estimated as $\phi_{dip} = (\phi_{1} + \phi_{4})/2$. The corresponding dip arrival time is given by $T_{dip} =T_{fold} +\phi_{dip}P_{0}$.\\
The values of the $\chi^{2}_{red}$ in Tab. \ref{tab:fit_par} are quite high. We can assume that this is a direct cause of the underestimation of the errors associated with the fitted data points. The uncertainties of all these data points can be increased by a constant factor to have a $\chi^{2}_{red}$ equal to 1, multiplying the uncertainties on the fitting parameters by a factor equal to the $\sqrt{\chi^{2}_{red}}$ of the fit.
As a consequence of this, for the calculations of the uncertainty on the fitting parameters, all the uncertainties relative to the values of $\phi_{dip} $ have been rescaled by the factor $\sqrt{\chi^{2}_{red}}$, when the best-fit model gave a $\sqrt{\chi^{2}_{red}}$ greater than 1. 
Moreover, we use a trial orbital period P$_{0}$ of 10590.864 s and a reference epoch T$_{0}$ of 10683.4663 TJD to obtain the delays of the dips' arrival times with respect to the reference epoch T$_{0}$. We choose the values of P$_{0}$ and T$_{0}$ arbitrarily, which are similar to those given by \cite{Levine}. 
The values of T$_{fold}$ are shown in Table \ref{tab:obs_log}, the best-fit parameters and the $\chi^{2}_{red}$ obtained by fitting the dips are shown in Table \ref{tab:fit_par}. Finally, the dip arrival times, the cycle  and the delays for each observation are shown in Table \ref{tab:arr_times}.\\
First, we  fit the obtained delays as a function of cycles with a linear function, taking into account only the errors associated with the delays. The fitting function is
\begin{equation}\label{eq:linear_law}
y(N)=a+bN,
\end{equation}
where N is the cycle number, $b$ is the correction to the trial orbital period ($\Delta P_{0}$) in seconds, and $a$ is the correction to the trial reference epoch ($\Delta T_{0}$) in seconds. 
Thus, we obtain a first estimation of the corrections to the reference epoch and to the orbital period.
Taking into account the error associated with the number of the cycles ($\Delta x$), we obtain a total error $\Delta_{tot} =\sqrt{(\Delta y)^{2} + (b*\Delta x)^{2}}$, where $\Delta y$ is the error associated with the dip arrival time \citep[see][]{Iaria}.
We repeated the fitting procedure, obtaining a $\chi^{2}(d.o.f.)$ of 37.81(10).
For the same reason adduced before, the uncertainties of the parameters returned by the linear fit have been rescaled by the factor $\sqrt{\chi^{2}_{red}}$, when the best-fit model gave a $\sqrt{\chi^{2}_{red}}$ greater than 1.
The best-fit model parameters are shown in Table \ref{tab:fit_res}. The orbital ephemeris obtained with the linear model is
\begin{equation}\label{eq:linear_ephem}
T_{dip}(N)= {\rm TJD(TDB)} \; 10683.4644(5) +  \frac{10590.896(2)}{86400} N,
\end{equation}
where 10683.4644(5) TJD and 10590.896(2) s are the reference epoch and orbital period, respectively. The associated errors are at 68\% confidence level. \\
We also try to fit the delays using a quadratic function:
\begin{equation}\label{eq:quad_law}
y(N) = a + bN + cN^{2},
\end{equation}
where $a$ is the reference epoch correction ($\Delta T_{0}$) in seconds, $b$ is the orbital period correction ($\Delta P_{0}$) in seconds, and $c=\frac{1}{2}P_{0}\dot{P}$ in units of seconds.
The best-fit model parameters are reported in Table \ref{tab:fit_par}. 
The fit returns a value of the $\chi^{2}(d.o.f.)$ of 36.04(9), while the F-test probability of chance improvement with respect to the previous linear fit is only of 52\%. This suggests that, adopting the quadratic ephemeris, we do not improve  the fit significantly. The orbital ephemeris obtained with the quadratic fit is
\begin{equation} \begin{split}\label{eq:quad_ephem}
T_{dip}(N)= {\rm TJD(TDB)} \; 10683.4641(7) +  \frac{10590.896(2)}{86400} N + \\
+ \frac{4(7)\times 10^{-8}}{86400} N^{2},
\end{split}\end{equation}
where 10683.4641(7) TJD is the new reference epoch, 10590.896(2) s is the new orbital period and $\dot{P}=(0.8 \pm 1.3)\times 10^{-11} s/s$ is the orbital period derivative obtained by the $c$ parameter that was returned by the fit. The associated errors are at 68\% confidence level. In the upper panel of Fig. \ref{fig:fit_trend} we show the delays versus the orbital cycles and, superimposed, two best-fit models, i.e. the linear and the quadratic ephemeris. At the bottom, the residuals resulting from the linear fit are shown.
The maximum discrepancy between the delays and the linear best-fit model is of about 300 s, which is about 2.8\% of the orbital period.

\begin{table}[h]
\caption{Journal of  arrival times of the X-ray dips obtained from each folded light curve. The number of dips in each grouped observation is also shown.} \label{tab:arr_times}         
     
\centering          
\small
\begin{tabular}{clccc}     
\hline\hline       

Point &
Dip time &
Cycle &
Delay  &
Number of \\

 &
(TJD;TDB) &
 &
(s) &
dips \\

\hline

1 &     
6108.5412(7) &
-37322 &
 -1303$\pm$64 &
 12\\

2 &     
9568.7220(7) &
-9094  & 
-592$\pm$56 &
2 \\

3 &     
10565.0554(10) &
-966 &
73$\pm$84 &
8 \\

4 &     
10683.4663(9) &
0 &
0 $\pm$79 &
11 \\

5 &     
11223.3058(7) &
 4404   &
-36$\pm$63 &
9 \\

6 &     
11577.8062(8) &
7296 &
20$\pm$70 &
7 \\

7 &     
12668.766(3) &
16196 &
254$\pm$216 &
5 \\

8 &     
12907.6725(9) &
18145 &
189$\pm$78  &
4 \\

9 &     
13370.2904(13) &
21919 &
451$\pm$110  &
6 \\

10 &            
14110.3063(9) &
27956 &
 779$\pm$76 &
 10 \\

11 &            
15601.9779(11)&
40125 &
988$\pm$95 &
7 \\

12 &            
15917.0112(8) &
42695 &
1339$\pm$65 &
14 \\

\hline

\hline
              
\end{tabular}

\end{table}



\begin{table}[h]
\caption{Best-fit values obtained from the linear and quadratic fits on the delays of the dips arrival times.} \label{tab:fit_res} 
     
\centering          
\small

\begin{tabular}{lcc}     
\hline\hline       

Parameter &
Linear &
Quadratic \\

\hline

a (s) &
-163 $\pm$ 23 &
-186 $\pm$ 29  \\

b ($\times10^{-3}$ s)&  
31.9 $\pm$ 0.9 &
32 $\pm$ 1      \\

c ($\times10^{-8}$ s) & 
- &
4 $\pm$ 7 \\

$\chi^{2}$(d.o.f) &     
 37.81(10) &
 36.04(9)       \\

\hline

\hline
              
\end{tabular}

\end{table}

To verify the goodness of the linear ephemeris shown in Eq. \ref{eq:linear_ephem}, we folded the 
RXTE/ASM light curve in the 2-10 keV energy band. This light curve covers a time interval of 15.5 years (from 5 January 1996 to 24 September 2011) and  barycentric corrections have been applied using the \verb|faxbary| tool.
We show the folded light curve using the reference epoch and orbital period suggested by \cite{Levine} and by our linear ephemeris in the left and right panels of Fig. \ref{fig:folded_comparison}, respectively. We adopted 14 phase bins per period (each phase bin corresponds to $\sim$757 s). We note that, using T$_{0}$ and P$_{orb}$ from Eq. \ref{eq:linear_ephem}, the dip occurs at phase zero, while using the values from \cite{Levine}, the dip occurs at a phase close to 0.1.

\section{Spectral analysis} 
\label{sec:spec}

We present the spectral analysis of 4U 1323-619 performed with \textit{INTEGRAL} data. This analysis is useful in Sect. \ref{sec:discussion} when extracting the unabsorbed flux of the source between 0.5 and 100 keV, to obtain an estimation of the luminosity.\\
We  found a number of \textit{INTEGRAL} pointings (science windows, SCW) collected on 2003 July 10--13 in the field of 4U 1323-619.
To maximize the spectral response, we selected only SCWs (with a typical duration of 3 ks)
with the source located within 4.5\degree and 3.5\degree from the centre of the IBIS and JEM-X FOVs, respectively.
Thus, we  analysed IBIS/ISGRI \citep{lebrun03} data of 80 SCWs and JEM-X2 \citep{lund03}
data of 51 SCWs (JEM-1 was switched-off at that time) with the standard INTEGRAL software OSA 10.1 \citep{courvoisier03}
We  verified that, in the JEM-X2 lightcurve,   
no type-I bursts were present and that the source flux was almost constant,
so that, an averaged spectra (3-35 keV) has been extracted in 16 channels for a total exposure time of 156 ks.
Because of the faintness of the source in hard X-rays,
the ISGRI spectrum has been extracted by the total (about 198 ks) mosaic image with the
\verb|mosaic_spec| tool in four energy bins, spanning from 22 keV up to 65 keV.

\begin{figure}
         \centering
         
         \includegraphics[angle=-90, width=9cm]{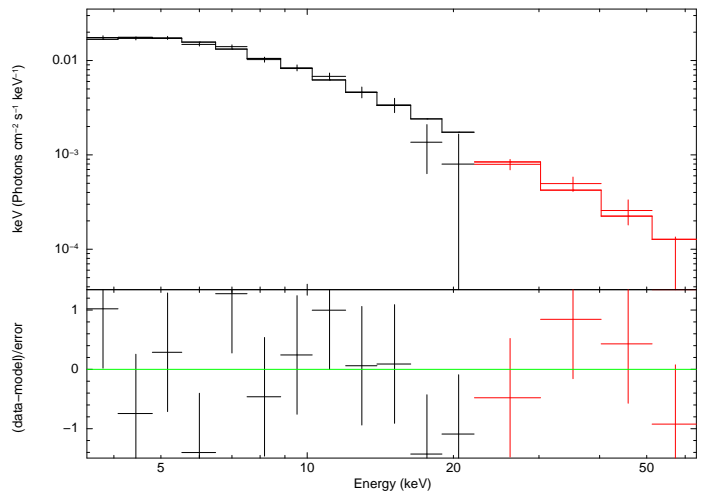}
  
         \caption{Upper panel: JEM-X (black) and ISGRI (red) spectra of 4U1323-619 in the energy range 3-65 keV.
  Lower panel: residuals in units of $\sigma$. }
         
         \label{fig:spec_model_JEMX/ISGRI}

         \end{figure}

The spectral analysis was  performed using XSPEC v. 12.8.2 \citep{Arnaud}.
We modelled the joint JEM-X2 and ISGRI spectrum with a thermal Comptonisation (\verb|nthComp| in XSPEC).  
This spectral component takes into account the nature of the seed photons (represented by the $inp\_type$ parameter that is equal to 0 for blackbody seed photons or to 1 for disk blackbody seed photons), the temperature kT$_{bb}$ of the seed photons, the electron temperature kT$_{e}$, an asymptotic power-law photon index $\Gamma$, the redshift, as well as a normalisation constant.
The absorption that is due to the interstellar medium has been taken into account by using the \verb|TBabs| model with the chemical abundances of \cite{Asplund} and the cross-sections of \cite{Verner}.
The energy domain of the spectrum, however, does not enable us to well constrain the value of N$_{H}$. As a consequence of this, we fix the column density of neutral hydrogen to the value found by \cite{Parmar}, which is also the highest available in literature. The fit gives a $\chi_{red}^2$ of 1.07 with 11 degrees of freedom. 
The best-fit parameters are shown in Tab. \ref{tab:fit_spec_res}, while the deconvolved model and the resulting residuals in unity of $\sigma$ are shown in Fig. \ref{fig:spec_model_JEMX/ISGRI}.

\begin{table}
\caption{Best-fit parameters of the spectral fitting on the JEM-X and ISGRI spectrum. \label{tab:fit_spec_res}}             
     
\centering          
\normalsize
\begin{tabular}{llc}    
\hline\hline       

Component &
Parameter &
Value \\

\hline

Tabs & 
N$_{H}$ ($\times10^{22} \rm cm^{-2}$) &
4 (frozen) \\

nthComp & 
$\Gamma$ & 
3.0$^{+0.4}_{-0.5}$ \\

nthComp  & 
kT$_{e}$ (keV)& 
>10 \\

nthComp  & 
kT$_{bb}$ (keV)& 
1.13$^{+0.11}_{-0.15}$\\

nthComp  & 
inp\_type & 
0 (frozen)  \\

nthComp  & 
Redshift & 
0 (frozen)  \\

nthComp  & 
Norm. ($\times10^{-3}$) & 
6.9$^{+1.9}_{-1.3}$ \\

%

\hline

  &
$\chi^{2}_{red}$(d.o.f.) &
1.069(11) \\

\hline

\end{tabular}

\tablefoot{The fitting parameters were set equal for both the JEM-X and ISGRI data-sets. The uncertainties are given at 90\% confidence level.} \\
\end{table}

\section{Discussion}
\label{sec:discussion}

\begin{figure}
         \centering
         
        \includegraphics[angle=-90, width=9cm]{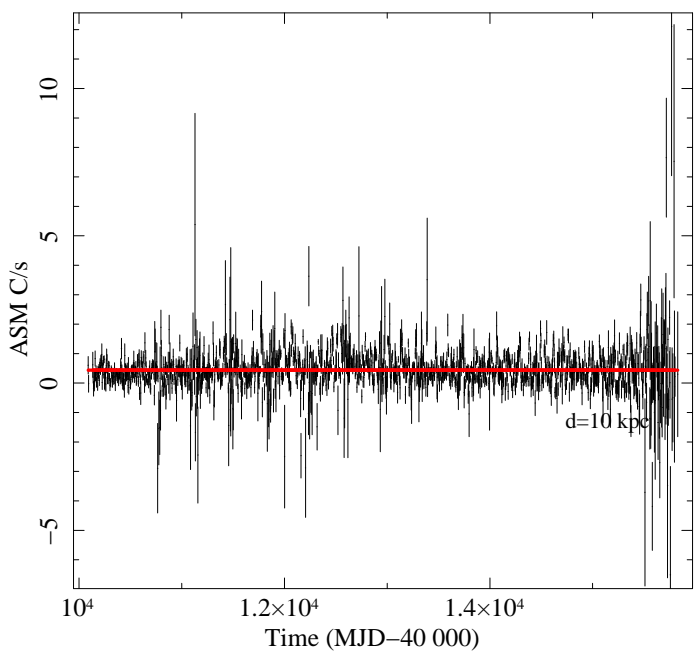}
  
         \caption{RXTE/ASM light curve in the 2-10 keV energy band. The red line represents the mean count rate maintained by 4U 1323-619 over 15.5 years. The bin time is  four days.}
         
         \label{fig:asm}
        \end{figure}

\begin{figure}
         \centering

        \includegraphics[angle=0, width=9cm]{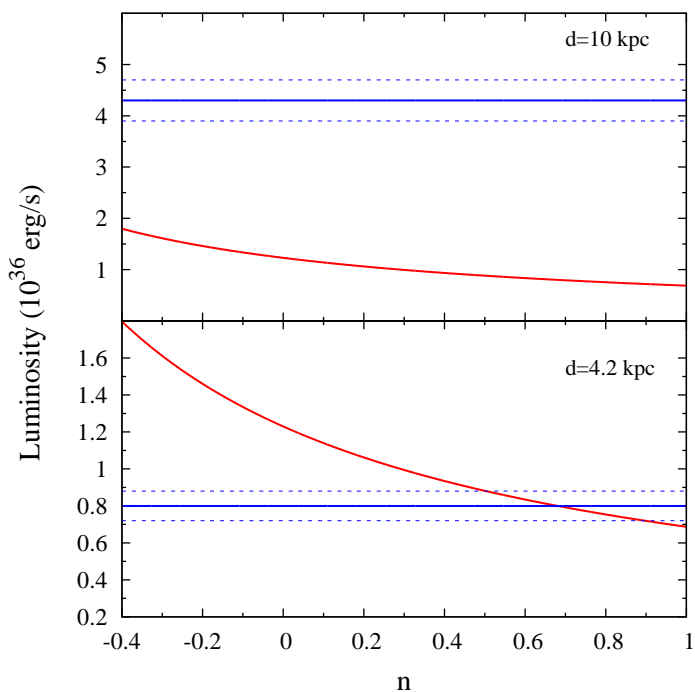}
  
         \caption{The red lines represent the luminosity as a function of the index $n$ for a distance of 10 kpc (upper panel) and 4.2 kpc (lower panel). The blue continuous lines represent the best values of the mean luminosity, while the dashed lines represent the associated errors.}
         
         \label{fig:L_10}
        \end{figure}

We estimated the orbital ephemeris of 4U 1323-619 using the whole available X-ray archival data from 1985 to 2011. We produced the linear and quadratic orbital ephemerides, finding that the value of the orbital period is 2.9419156(6) hrs. This is compatible with  previous estimations \citep[see][]{Levine}, but with an accuracy that is ten times higher than that proposed by \cite{Levine}. \\
In a binary system the accretion of matter from the companion star onto the neutron star is the main mechanism of generation of radiation. The mass accretion rate, however, is governed by the angular momentum losses to which the binary system is subjected. 
The main channels of angular momentum loss are: the mass loss from the binary system as a consequence of the accretion, the generation of gravitational waves (especially in close binaries composed by massive stars), and the magnetic braking, which consists of a transfer of angular momentum from the companion star to the ionised matter that surrounds it and that interacts with its magnetic field.  
The redistribution of the angular momentum induced by these interactions causes a variation of the orbital period of the binary system.
As a consequence of this, an estimation of the orbital period derivative is important to understand the orbital evolution of the system.\\
As a first step towards the extrapolation of an evolutive scenario for 4U 1323-619, we infer the companion star mass M$_{2}$.  Assuming that 4U 1323-619 is a persistent X-ray source in a Roche lobe overflow regime, as suggested by the long time monitoring of the RXTE/ASM (see Fig. \ref{fig:asm}), then the companion star fills its Roche lobe and, consequently, the companion star radius R$_{2}$ is equal to the Roche lobe radius R$_{L2}$. \\
The Roche lobe radius is given by the expression
\begin{equation}\label{eq:Pacinzsky}
R_{L2}=0.46224\;a\left(\frac{m_{2}}{m_{1}+m_{2}}\right)^{1/3}
\end{equation}
of \cite{Pac}, where $m_{1}$ and $m_{2}$ are the NS and companion star masses in units of solar masses and $a$ is the orbital separation of the binary system.
Assuming that the companion star belongs to the lower main sequence, we adopt the relation 
\begin{equation}\label{eq:m2_th}
\frac{R_{2}}{R_{\odot}}=0.877\;m^{0.807}_{2}
,\end{equation}
valid for M-stars \citep{Neece}. Combining equations \ref{eq:Pacinzsky} and \ref{eq:m2_th} with the third Kepler law 
and assuming a neutron star (NS) mass of M$_{1}$=1.4 M$_{\odot}$ and the orbital period obtained from our ephemeris, we find that the companion star mass is M$_{2}\simeq$0.28$\pm$0.03 M$_{\odot}$. Here we took into account an accuracy of 10\% in the mass estimation \citep[see][]{Neece}. 
Our result is compatible with 0.25 M$_{\odot}$, as suggested by \cite{Zolotukhin}.
Since the companion star mass is 0.28 M$_{\odot}$ we expect that it is fully convective and, consequently, it cannot have a magnetic field anchored to it \citep{Nelson}. This implies that  magnetic braking as a driving mechanism of the orbital evolution is ruled out.
Assuming that the unique mechanism of angular momentum loss in this binary system is due to the gravitational radiation and, considering a conservative mass transfer, we can infer the secular mass accretion rate by re-arranging Eq. 10 of \cite{Burderi}, i.e.
\begin{equation}\label{eq:m_dot}
\dot{m}_{-8}=0.6\;\frac{1}{3q-\frac{3}{2}n-\frac{5}{2}}\;m^{8/3}_{1}q^{2}(1+q)^{-1/3}P^{-8/3}_{2h},
\end{equation}
where $\dot{m}_{-8}$ is the mass transfer rate from the companion star in units of $10^{-8}$ M$_{\odot}$ yr$^{-1}$, $q$ is the ratio of m$_{2}$ to m$_{1}$, and P$_{2h}$ is the orbital period in units of two hours. The index $n$ can assume the values -1/3 or 0.8 and  is associated with the internal structure of the companion star. If the star is in thermal equilibrium, its index $n$ is 0.8 (the index of the mass-radius relation we adopted in Eq. \ref{eq:m2_th}) whilst it is equal to the adiabatic index $n= -1/3$ otherwise.
The values of $\dot{m}$ predicted by Eq. \ref{eq:m_dot} are 6.4$\times10^{-11}$ M$_{\odot}$ yr$^{-1}$ for $n$=0.8 and 1.42$\times10^{-10}$ M$_{\odot}$ yr$^{-1}$ for $n$=-1/3.\\
Following the theory of secular evolution for X-ray binary systems, we can write the X-ray luminosity as 
 \begin{equation} \begin{split} \label{eq:lum}
L=\frac{GM_{1}\dot{M}}{R_{1}}= &\\ 5.0\times10^{37}\frac{1}{3q-\frac{3}{2}n-\frac{5}{2}}\;m^{11/3}_{1}q^{2}(1+q)^{-1/3}P^{-8/3}_{2h} {\rm erg\;s^{-1}}
\end{split}
\end{equation}
\citep[see][]{King}, where we substitute for $\dot{M}$ the expression of Eq. \ref{eq:m_dot}, and assume an NS radius $R_{1}$ of 10 km. 
Since we need to compare the predicted luminosity with the bolometric observed luminosity, we extrapolate the 0.5-100 keV luminosity from all the available observations.
For some observations the spectral analysis of the source was already shown in literature.
Since the energy spectrum in this case is dominated by a cut-off power-law component with a cut-off energy at 85 keV \citep{Belucinska}, we only use  this spectral component to extrapolate the unabsorbed luminosity, using the photon index and normalisation values reported in literature for each observation and fixing the cut-off energy at 85 keV. We find a flux of 3.06$\times10^{-10}$, 3.61$\times10^{-10}$, 3.38$\times10^{-10}$, and 3.62$\times10^{-10}$ erg cm$^{2}$ s$^{-1}$ for the observations taken by \textit{Suzaku} \citep{Belucinska}, \textit{EXOSAT} \citep{Parmar}, \textit{BeppoSAX} \citep{Belucinska_beppo}, and \textit{RXTE} \citep{Galloway}, respectively.
On the other hand, we extract a flux of 4.7$\times10^{-10}$ erg cm$^{2}$ s$^{-1}$ in the 0.5-100 keV energy band from the \textit{XMM-Newton} observation, and of 3.3$\times10^{-10}$ erg cm$^{2}$ s$^{-1}$ from the JEM-X/ISGRI observation. \\
If we consider the 2-10 keV RXTE/ASM light curve spanning 15.5 years of data, from 3 March 1996 (10145 TJD) to 12 September 2011 (15816 TJD), we can observe that the source maintained a roughly constant count-rate (see Fig. \ref{fig:asm}). Fitting the ASM light curve with a constant function, in fact, we obtain a count rate of (0.483$\pm$0.011) c s$^{-1}$ with a $\chi^{2}(d.o.f.)=$5042(1388).\\
As a consequence of this, we estimate the mean flux for 4U 1323-619, averaging all the values of flux just obtained. The mean value of flux is of 3.6$\times10^{-10}$ erg cm$^{2}$ s$^{-1}$. 
To be more conservative, we verified that this result is not sensitive to the specific value of N$_{H}$ that was previously used for the JEM-X/ISGRI fit, which is the maximum available in literature. Adopting the minimum value of N$_{H}$ available in literature \citep[3.2$\times10^{22}$ cm$^{-2}$;][]{Belucinska}, we extract a flux of 3.2$\times10^{-10}$ erg cm$^{2}$ s$^{-1}$ for the JEM-X/ISGRI observation in the 0.5-100 keV energy band. Again, we obtain a mean value of flux of 3.6$\times10^{-10}$ erg cm$^{2}$ s$^{-1}$ averaging all the values of flux obtained above.\\
Using the mean flux and adopting a distance of 10 kpc we determine a mean luminosity of (4.3 $\pm$ 0.4)$\times 10^{36}$ erg s$^{-1}$ for 4U 1323-619.
In this estimation, we took into account an arbitrary error of 10\%.
We observe that the luminosity predicted by Eq. \ref{eq:lum} can reach a maximum value of 1.7$\times10^{36}$ erg s$^{-1}$ for n=-1/3, and a value of 7.5$\times10^{35}$ erg s$^{-1}$ for n=0.8.
We show the mean luminosity together with that predicted by Eq. \ref{eq:lum} in the upper panel of Fig. \ref{fig:L_10}.
From this comparison, we can conclude that, for a distance of 10 kpc, the observed mean luminosity is not in accordance with that predicted by the theory of the secular evolution. 
We note that a wrong distance value could contribute to the observed inconsistency.

\begin{figure}
         \centering
         
        \includegraphics[angle=-90, width=9cm]{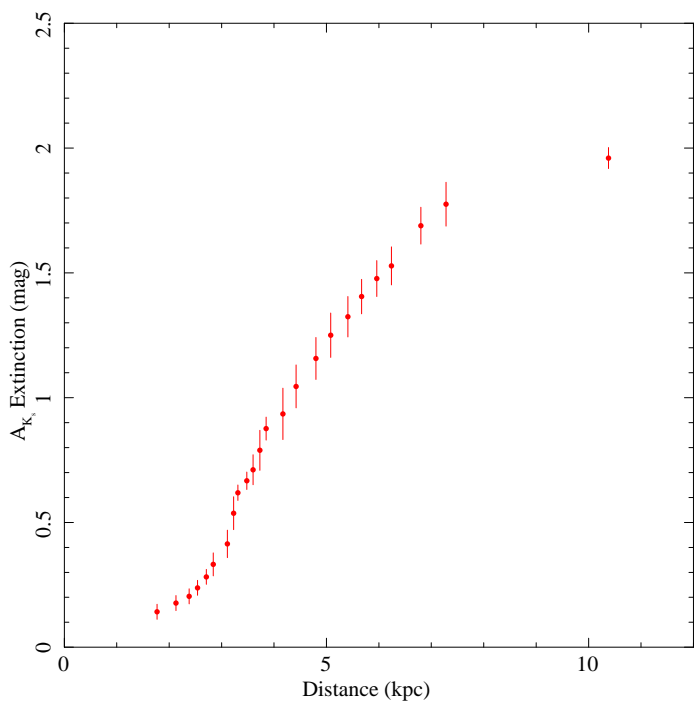}
  
         \caption{Extinction of the K$_{s}$ radiation in the direction of the source ($l=307\degree$ , $b=0.5\degree$) as a function of the distance \citep{Marshall}.}
         
         \label{fig:extinction}
        \end{figure}

For this purpose, we take into account the 3D extinction map of the radiation in the  K$_{s}$ band for our Galaxy \citep{Marshall}. This map suggests how great  the extinction of the radiation is in the direction of the source ($l=307\degree$, $b=0.5\degree$) as a function of the distance. In Fig. \ref{fig:extinction} we report the extinction predicted by the model of \cite{Marshall} as  a function of the distance from the source. We extrapolated the profile, taking into account the galactic coordinates of 4U 1323-619. We calculate the visual extinction using the \cite{Guver} relation
\begin{equation}\label{eq:A_v}
N_{H}=(2.21 \pm 0.09)\times10^{21}A_{V},
\end{equation}
where N$_{H}$ is the equivalent column density of neutral hydrogen of the absorbing interstellar matter and A$_{V}$ is the visual extinction of the source radiation. Using the N$_{H}$ value 3.2$\times10^{22}$ cm$^{-2}$ found by \cite{Belucinska}, we estimate a visual extinction of A$_{V}$=14.5$\pm$0.7 mag that is compatible with the value found by \cite{Belucinska_beppo}. 
The visual extinction is related to the extinction of the radiation in the K$_{s}$ band through the relation
\begin{equation}\label{eq:A_ks}
A_{K_{S}}=(0.062\pm0.005)\;A_{V}\;{\rm mag}
\end{equation}
\citep{Nishiyama}, where A$_{K_{S}}$ is the extinction in the K$_{S}$ band. With this relation we find a value for the extinction of A$_{K_{S}}$= 0.90$\pm$0.09 mag.\\
We fitted the profile of the extinction of Fig. \ref{fig:extinction} with a quadratic function between 3 and 6 kpc, imposing a null value of the extinction for a null distance. The fit gives us a $\chi^{2}_{red}$ of 1.38 with 12 degrees of freedom, a linear parameter of 0.10$\pm$ 0.02, and a quadratic parameter of 0.028$\pm$0.004. The uncertainties are given with a confidence level of 68\%. According to the value of A$_{K_{S}}$, obtained before, we estimate a distance of 4.2$^{+0.8}_{-0.7}$ kpc with a confidence level of 68\%. This distance is fully in agreement with the value of distance suggested by \cite{Zolotukhin}. \\
Taking into account a distance of 4.2$_{-0.7}^{+0.8}$ kpc, we infer a luminosity of (0.8 $\pm$ 0.3)$\times 10^{36}$ erg s$^{-1}$ that is compatible with the one predicted by Eq. \ref{eq:lum} for a $n$ index close to 0.8 and thus for a thermal equilibrium state of the companion star (see Fig. \ref{fig:L_10}). \\
A value of luminosity that completely matches the one predicted by Eq. \ref{eq:lum} for $n$=0.8 is obtained for a distance to the source of d$\sim$4.17 kpc. According to the \cite{Marshall} model (see Fig. \ref{fig:extinction}), an extinction of 0.88 mag competes for this value of distance in the K$_{s}$ band, and thus, by Eq. \ref{eq:A_ks} and Eq. \ref{eq:A_v}, we infer A$_{v}$=14.2 mag and N$_{H}\sim$3.1$\times10^{22}$ cm$^{-2}$. This value of N$_{H}$ is fully in agreement with that obtained by \cite{Belucinska}.\\
We also explored a non-conservative mass transfer scenario. We observe that in the case in which the mass is expelled from the position of the companion star, for an index n=0.8 and a distance from the source of 4.2 kpc, the luminosities obtained from the individual X-ray observations are compatible with the luminosity predicted by a non-conservative mass transfer theory for a value of $\beta \geq$ 0.8, where $\beta$ is the fraction of matter accreted onto the neutron star. This implies that the observations are in agreement with a scenario in which more than the 80\% of the mass is accreted onto the neutron star. Thus, on the basis of the fact that only the 20\% of mass of the companion star is lost from the system, we can confirm that the mass accretion mechanism could be rightly considered conservative. \\
To investigate  the orbital evolution of the system, we need to compare the Kelvin-Helmholtz time-scale $\tau_{KH}$ (i.e. the characteristic time to reach the thermal equilibrium for a star) with the mass transfer time-scale $\tau_{\dot{M}}$. We calculate the Kelvin-Helmholtz time-scale adopting the relation
\begin{equation}\label{eq:t_m}
\tau_{KH}=3.1\times10^{7}\left(\frac{M_{2}}{M_{\odot}}\right)^{2}\frac{R_{\odot}}{R_{2}}\frac{L_{\odot}}{L}
\end{equation}
of \cite{Verbunt}, together with the mass-luminosity relation for a M-type star 
\begin{equation}\label{eq:m_L}
\frac{L_{2}}{L_{\odot}}=0.231\;\left(\frac{M_{2}}{M_{\odot}}\right)^{2.61}
\end{equation}
of \cite{Neece}. Moreover, we assume the \cite{Neece} mass-radius relation of Eq. \ref{eq:m2_th} for a main sequence M-type star in equilibrium. Our estimation of the Kelvin-Helmholtz time-scale is of $\tau_{KH}$=9$\times10^{8}$ yrs.\\
On the other hand, we extrapolate the mass transfer time-scale using the relation
\begin{equation}\label{eq:tau_m}
\tau_{\dot{M}}=\frac{m_{2}}{\dot{m}}=\frac{G\;m_{1} m_{2}}{L\;R_{NS}},
\end{equation}
where $L$ is the source luminosity. To achieve this, we use the mean luminosity estimated above during the \textit{ASM} monitoring, obtaining a mass transfer time-scale of 4.4$\times10^{9}$ yrs. This value is greater than the Kelvin-Helmholtz time-scale by a factor of 4.9.
This implies that the companion star is in thermal equilibrium and we can conclude that the mass-radius relation in Eq. \ref{eq:m2_th} has been correctly used, suggesting that the proposed scenario is self-consistent. \\
To understand how the binary system is going to evolve, we estimate the orbital period derivative re-arranging Eq. 4 of \cite{Burderi}:
\begin{equation}\label{eq:P_dot_Burderi}
\dot{m}_{-8}=87.5\;(3n-1)^{-1}m_{2}\left(\frac{\dot{P}_{-10}}{P_{2h}}\right),
\end{equation}
where $\dot{P}_{-10}$ is the orbital period derivative in units of 10$^{-10}$ s/s and $P_{2h}$ is the orbital period of the system in units of two hours. Adopting an index n=0.8, we obtain a value of the period derivative of $\dot{P}\sim-5.4\times10^{-14}$ s/s that is compatible with the constraint on the $\dot{P, }$ which is estimated from our ephemeris, i.e. $(0.8\pm1.3)\times10^{-11}$ s/s. From this result, we can infer that the orbital period is decreasing and that the system is shrinking as a consequence of  the orbits being Keplerian.\\

\section{Conclusions}

We constrain the X-ray source position of 4U 1323-619, finding out that the source is located inside a circular area centred at RA (J2000)= 201.6543\degree and DEC (J2000)= -62.1358\degree  and with a radius of 0.0002\degree (that is 0.6"). This result allows us to confirm the suggestion of \cite{Zolotukhin} that the B source in Fig. \ref{fig:source_position} is the IR counterpart of 4U 1323-619.\\
In addition, using observations from 1985 to 2011, we infer for the first time the linear and quadratic orbital ephemerides for 4U 1323-619. We estimate the orbital period of the binary system with an accuracy ten times higher than that which was proposed by \cite{Levine}. We obtain a refined measure of the period of $P= 2.9419156(6) $ hrs, in line with previous estimates, as reported in the literature. We infer for the first time a weak constraint on the orbital period derivative of the system that is of $\dot{P}= (8\pm 13)\times10^{-12}$ s/s. Assuming a fully conservative mass-transfer scenario and that the companion star is an M-type main-sequence star, we estimate the mass of the companion to be of 0.28$\pm$0.03 M$_{\odot}$. This result suggests that the star is fully convective and that the magnetic braking mechanism can be ruled out as an explanation of angular momentum losses from the binary system, which is therefore driven by the mechanism of gravitational radiation. \\ 
We inferred that the companion star transfers matter onto the neutron star surface via the inner Lagrangian point in a conservative regime. 
According to the conservative mass-transfer scenario, and taking into account the map of the K$_{s}$ radiation extinction in our Galaxy, we estimate a distance of (4.2$^{+0.8}_{-0.7}$) kpc at 68\% confidence level.

\section*{Acknowledgements}

This research has made use of data and/or software provided by the High Energy Astrophysics Science Archive Research Center (HEASARC), which is a service of the Astrophysics Science Division at NASA/GSFC and the High Energy Astrophysics Division of the Smithsonian Astrophysical Observatory.
This research  made use of the VizieR catalogue access tool, CDS, Strasbourg, France.
The High-Energy Astrophysics Group of Palermo acknowledges support
from the Fondo Finalizzato alla Ricerca (FFR) 2012/13, Project
N. 2012-ATE-0390, founded by the University of Palermo.
This work was partially supported by the Regione Autonoma della
Sardegna through POR-FSE Sardegna 2007-2013, L.R. 7/2007,
Progetti di Ricerca di Base e Orientata, Project N. CRP-60529.
We also acknowledge financial contributions from the agreement
ASI-INAF I/037/12/0. AR acknowledges Sardinia Regional Government
for the financial support (P.O.R. Sardegna F.S.E. Operational
Programme of the Autonomous Region of Sardinia, European Social
Fund 2007-2013 - Axis IV Human Resources, Objective l.3, Line of
Activity l.3.1.).
MDS thanks the Dipartimento di Fisica e Chimica, Università di Palermo,
for their hospitality.


\bibliographystyle{aa} 
\bibliography{4u1323_619}

\begin{thebibliography}{30}
\expandafter\ifx\csname natexlab\endcsname\relax\def\natexlab#1{#1}\fi

\bibitem[{{Arnaud}(1996)}]{Arnaud}
{Arnaud}, K.~A. 1996, in Astronomical Society of the Pacific Conference Series,
  Vol. 101, Astronomical Data Analysis Software and Systems V, ed. G.~H.
  {Jacoby} \& J.~{Barnes}, 17

\bibitem[{{Asplund} {et~al.}(2009){Asplund}, {Grevesse}, {Sauval}, \&
  {Scott}}]{Asplund}
{Asplund}, M., {Grevesse}, N., {Sauval}, A.~J., \& {Scott}, P. 2009, \araa, 47,
  481

\bibitem[{{Ba{\l}uci{\'n}ska-Church} {et~al.}(1999){Ba{\l}uci{\'n}ska-Church},
  {Church}, {Oosterbroek}, {Segreto}, {Morley}, \& {Parmar}}]{Belucinska_beppo}
{Ba{\l}uci{\'n}ska-Church}, M., {Church}, M.~J., {Oosterbroek}, T., {et~al.}
  1999, \aap, 349, 495

\bibitem[{{Ba{\l}uci{\'n}ska-Church} {et~al.}(2009){Ba{\l}uci{\'n}ska-Church},
  {Dotani}, {Hirotsu}, \& {Church}}]{Belucinska}
{Ba{\l}uci{\'n}ska-Church}, M., {Dotani}, T., {Hirotsu}, T., \& {Church}, M.~J.
  2009, \aap, 500, 873

\bibitem[{{Boirin} {et~al.}(2005){Boirin}, {M{\'e}ndez}, {D{\'{\i}}az Trigo},
  {Parmar}, \& {Kaastra}}]{Boirin}
{Boirin}, L., {M{\'e}ndez}, M., {D{\'{\i}}az Trigo}, M., {Parmar}, A.~N., \&
  {Kaastra}, J.~S. 2005, \aap, 436, 195

\bibitem[{{Burderi} {et~al.}(2010){Burderi}, {Di Salvo}, {Riggio}, {Papitto},
  {Iaria}, {D'A{\`i}}, \& {Menna}}]{Burderi}
{Burderi}, L., {Di Salvo}, T., {Riggio}, A., {et~al.} 2010, \aap, 515, A44

\bibitem[{{Courvoisier} {et~al.}(2003){Courvoisier}, {Walter}, {Beckmann},
  {Dean}, {Dubath}, {Hudec}, {Kretschmar}, {Mereghetti}, {Montmerle},
  {Mowlavi}, {Paltani}, {Preite Martinez}, {Produit}, {Staubert}, {Strong},
  {Swings}, {Westergaard}, {White}, {Winkler}, \& {Zdziarski}}]{courvoisier03}
{Courvoisier}, T.~J.-L., {Walter}, R., {Beckmann}, V., {et~al.} 2003, \aap,
  411, L53

\bibitem[{{Forman} {et~al.}(1978){Forman}, {Jones}, {Cominsky}, {Julien},
  {Murray}, {Peters}, {Tananbaum}, \& {Giacconi}}]{Forman}
{Forman}, W., {Jones}, C., {Cominsky}, L., {et~al.} 1978, \apjs, 38, 357

\bibitem[{{Frank} {et~al.}(1987){Frank}, {King}, \& {Lasota}}]{Frank}
{Frank}, J., {King}, A.~R., \& {Lasota}, J.-P. 1987, \aap, 178, 137

\bibitem[{{Galloway} {et~al.}(2008){Galloway}, {Muno}, {Hartman}, {Psaltis}, \&
  {Chakrabarty}}]{Galloway}
{Galloway}, D.~K., {Muno}, M.~P., {Hartman}, J.~M., {Psaltis}, D., \&
  {Chakrabarty}, D. 2008, \apjs, 179, 360

\bibitem[{{G{\"u}ver} \& {{\"O}zel}(2009)}]{Guver}
{G{\"u}ver}, T. \& {{\"O}zel}, F. 2009, \mnras, 400, 2050

\bibitem[{{Iaria} {et~al.}(2014){Iaria}, {Di Salvo}, {Burderi}, {Riggio},
  {D'A{\`i}}, \& {Robba}}]{Iaria}
{Iaria}, R., {Di Salvo}, T., {Burderi}, L., {et~al.} 2014, \aap, 561, A99

\bibitem[{{Iaria} {et~al.}(2015){Iaria}, {Di Salvo}, {Gambino}, {Del Santo},
  {Romano}, {Matranga}, {Galiano}, {Scarano}, {Riggio}, {Sanna}, {Pintore}, \&
  {Burderi}}]{Iaria_2015}
{Iaria}, R., {Di Salvo}, T., {Gambino}, A.~F., {et~al.} 2015, \aap, 582, A32

\bibitem[{{King}(1988)}]{King}
{King}, A.~R. 1988, \qjras, 29, 1

\bibitem[{{Lebrun} {et~al.}(2003){Lebrun}, {Leray}, {Lavocat}, {Cr{\'e}tolle},
  {Arqu{\`e}s}, {Blondel}, {Bonnin}, {Bou{\`e}re}, {Cara}, {Chaleil}, {Daly},
  {Desages}, {Dzitko}, {Horeau}, {Laurent}, {Limousin}, {Mathy}, {Mauguen},
  {Meignier}, {Molini{\'e}}, {Poindron}, {Rouger}, {Sauvageon}, \&
  {Tourrette}}]{lebrun03}
{Lebrun}, F., {Leray}, J.~P., {Lavocat}, P., {et~al.} 2003, \aap, 411, L141

\bibitem[{{Levine} {et~al.}(2011){Levine}, {Bradt}, {Chakrabarty}, {Corbet}, \&
  {Harris}}]{Levine}
{Levine}, A.~M., {Bradt}, H.~V., {Chakrabarty}, D., {Corbet}, R.~H.~D., \&
  {Harris}, R.~J. 2011, \apjs, 196, 6

\bibitem[{{Lund} {et~al.}(2003){Lund}, {Budtz-J{\o}rgensen}, {Westergaard},
  {Brandt}, {Rasmussen}, {Hornstrup}, {Oxborrow}, {Chenevez}, {Jensen},
  {Laursen}, {Andersen}, {Mogensen}, {Rasmussen}, {Om{\o}}, {Pedersen},
  {Polny}, {Andersson}, {Andersson}, {K{\"a}m{\"a}r{\"a}inen}, {Vilhu},
  {Huovelin}, {Maisala}, {Morawski}, {Juchnikowski}, {Costa}, {Feroci},
  {Rubini}, {Rapisarda}, {Morelli}, {Carassiti}, {Frontera}, {Pelliciari},
  {Loffredo}, {Mart{\'{\i}}nez N{\'u}{\~n}ez}, {Reglero}, {Velasco}, {Larsson},
  {Svensson}, {Zdziarski}, {Castro-Tirado}, {Attina}, {Goria}, {Giulianelli},
  {Cordero}, {Rezazad}, {Schmidt}, {Carli}, {Gomez}, {Jensen}, {Sarri},
  {Tiemon}, {Orr}, {Much}, {Kretschmar}, \& {Schnopper}}]{lund03}
{Lund}, N., {Budtz-J{\o}rgensen}, C., {Westergaard}, N.~J., {et~al.} 2003,
  \aap, 411, L231

\bibitem[{{Marshall} {et~al.}(2006){Marshall}, {Robin}, {Reyl{\'e}},
  {Schultheis}, \& {Picaud}}]{Marshall}
{Marshall}, D.~J., {Robin}, A.~C., {Reyl{\'e}}, C., {Schultheis}, M., \&
  {Picaud}, S. 2006, \aap, 453, 635

\bibitem[{{Neece}(1984)}]{Neece}
{Neece}, G.~D. 1984, \apj, 277, 738

\bibitem[{{Nelson} \& {Rappaport}(2003)}]{Nelson}
{Nelson}, L.~A. \& {Rappaport}, S. 2003, \apj, 598, 431

\bibitem[{{Nishiyama} {et~al.}(2008){Nishiyama}, {Nagata}, {Tamura}, {Kandori},
  {Hatano}, {Sato}, \& {Sugitani}}]{Nishiyama}
{Nishiyama}, S., {Nagata}, T., {Tamura}, M., {et~al.} 2008, \apj, 680, 1174

\bibitem[{{Paczy{\'n}ski}(1971)}]{Pac}
{Paczy{\'n}ski}, B. 1971, \araa, 9, 183

\bibitem[{{Parmar} {et~al.}(1989){Parmar}, {Gottwald}, {van der Klis}, \& {van
  Paradijs}}]{Parmar}
{Parmar}, A.~N., {Gottwald}, M., {van der Klis}, M., \& {van Paradijs}, J.
  1989, \apj, 338, 1024

\bibitem[{{Predehl} \& {Schmitt}(1995)}]{Predel}
{Predehl}, P. \& {Schmitt}, J.~H.~M.~M. 1995, \aap, 293, 889

\bibitem[{{Smale}(1995)}]{Smale}
{Smale}, A.~P. 1995, \aj, 110, 1292

\bibitem[{{van der Klis} {et~al.}(1985){van der Klis}, {Jansen}, {van
  Paradijs}, \& {Stollman}}]{Van_der_klis}
{van der Klis}, M., {Jansen}, F., {van Paradijs}, J., \& {Stollman}, G. 1985,
  \ssr, 40, 287

\bibitem[{{Verbunt}(1993)}]{Verbunt}
{Verbunt}, F. 1993, \araa, 31, 93

\bibitem[{{Verner} {et~al.}(1996){Verner}, {Ferland}, {Korista}, \&
  {Yakovlev}}]{Verner}
{Verner}, D.~A., {Ferland}, G.~J., {Korista}, K.~T., \& {Yakovlev}, D.~G. 1996,
  \apj, 465, 487

\bibitem[{{Warwick} {et~al.}(1981){Warwick}, {Marshall}, {Fraser}, {Watson},
  {Lawrence}, {Page}, {Pounds}, {Ricketts}, {Sims}, \& {Smith}}]{Warwick}
{Warwick}, R.~S., {Marshall}, N., {Fraser}, G.~W., {et~al.} 1981, \mnras, 197,
  865

\bibitem[{{Zolotukhin} {et~al.}(2010){Zolotukhin}, {Revnivtsev}, \&
  {Shakura}}]{Zolotukhin}
{Zolotukhin}, I.~Y., {Revnivtsev}, M.~G., \& {Shakura}, N.~I. 2010, \mnras,
  401, L1

\end{thebibliography}

\end{document}